# Resilience in Optical Wireless Systems

Sarah O. M. Saeed[1], Sanaa Hamid Mohamed[1], Osama Zwaid Alsulami[1], Mohammed T. Alresheedi[2], Taisir E. H. Elgorashi[1], and Jaafar M. H. Elmirghani[1]

[1]School of Electronic and Electrical Engineering, University of Leeds, LS2 9JT, U.K.
[2]Department of Electrical Engineering, King Saud University, Riyadh, Kingdom of Saudi Arabia
elsoms@leeds.ac.uk, elshm@leeds.ac.uk, ml15ozma@leeds.ac.uk, malresheedi@ksu.edu.sa, t.e.h.elgorashi@leeds.ac.uk, j.m.h.elmirghani@leeds.ac.uk

**ABSTRACT**
High reliability and availability of communication services is a key requirement that needs to be ensured by service providers. Since the direct line-of-sight (LOS) beam is prone to blockage in indoor optical wireless communication systems, a backup link needs to be at hand in case of blockage, and hence channel allocation algorithms should be blockage-aware. In this paper, the impact of beam blockage due to a disc with varying size and distance from the receiver is studied where blockage is quantitatively evaluated using percentage blockage for 512 room locations at 25 cm separation. It was found that assigning two links with maximum separation between the serving access points can reduce or eliminate blockage compared to the case when resilience is not implemented. Increasing the number of allocated access points per user further increases resilience.

## 1. INTRODUCTION

The data-intensive applications resulting from the increasing popularity of smart devices in addition to the trend of connecting everything to the Internet, requires exploration of new spectrums other than the already congested radio frequency (RF) spectrum. Optical Wireless (OW) communication is a candidate to fulfil these demands with the abundant unregulated spectrum which provides four orders of magnitude more bandwidth compared to the available RF bandwidth [1]. OW communication utilizes the Infra-red, visible, and ultra-violet optical spectrum. It is characterized by the use of inexpensive transmitters and receivers, in addition, since light does not penetrate opaque objects, this enables frequency reuse resulting in better spatial spectrum efficiency and better physical layer security. However, the inability of light to penetrate walls or opaque objects causes the loss of communication channel if an objects obstructs the LOS links. OWC links can also exploit the reflections of optical power from the different surfaces within a room, however, substantial amount of power is lost due to absorption and diffusion and also the multiple paths result in increased delay spread that causes Inter-Symbol-Interference (ISI) which decreases the achievable data rate. LOS links are characterized by better power efficiency and minimum delay spread which enable high data rate communication. However, LOS links are susceptible to beam blockage and hence resilience should be implemented to ensure the reliability of the communication service [2], [3]. The reported multi-gigabits per second systems in literature depend on engineering the OWC system by utilising multispot diffusing systems, diversity transmission and reception [4]–[6], and adaptation techniques with beam steering and Computer Generated Holograms [7]. The design of these systems results in improved Signal-to-Noise (SNR) ratio and reduced delay spread and hence higher data rates.
Resilience can be implemented by using backup optical links or backup radio links since OWC complements RF communication with the advantage of better mobility when moving between compartments. In this paper, we will follow the first approach of using optical protection links. We study resilience in a visible light communication (VLC) system where usually multiple luminaires exist and multiple connections between the user and the access points usually exist. The effect of blockage is studied using the same procedure as [8] where an opaque disc with parameters of radius, height, and horizontal distance from the receiver is assumed to block the LOS beam. Resilience with 2, 4, and 8 access points serving the user is studied using the percentage blockage as a metric. The rest of this paper is organized as follows: Section 2 describes the system model. Section 3 presents the results and discussion. Section 4 presents the conclusion and future work.

## 2. SYSTEM MODEL

An empty room with dimensions of 4 m, 8 m, and 3 m for the width, length, and height respectively is assumed. Receivers (users) are distributed over a so called; communication floor (CF) that is assumed to be at 1 m above the ground and this represents the typical desk height. Lighting is provided by eight sources of optical power which are used at the same time for communication as access points (APs). The APs are assumed at the coordinates of (1,1,3), (1,3,3), (1,5,3), (1,7,3), (3,1,3), (3,3,3), (3,5,3), and (3,7,3) as shown in Fig.1. A controller, which is assumed to have perfect knowledge of the channel state, allocates the access points to users [9]. However, the controller allocates multiple links to mitigate blockage.
To study blockage, a simplified disc object is assumed to obstruct the LOS link. More complexity can be added to represent real objects in the indoor environment, for example this can be changed to a column for the human body as in [10], however, this is planned for future and for the scope of this paper only the disc is considered. The disc has three parameters which are radius, height, and horizontal separation from the receiver in the positive y-direction.

In this study, two parameters of the disc are fixed while the third is varied. For a given set of parameters and a given APs allocation, the receiver blockage state is studied over the whole CF with 25 cm separation, i.e., 512 locations for the room considered. The percentage blockage is calculated as the percentage of the number of locations where the LOS beam is blocked to the total number of locations. Three cases were studied that are selected based on the number of allocated APs and their separation; assuming 2 assigned APs, 4 APs and 8 APs. These can be compared with the reference case in [8].

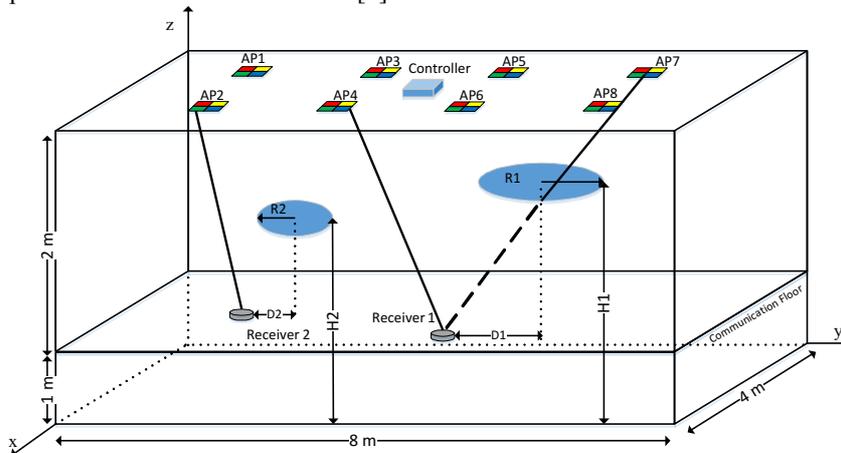

*Figure 1. Room model with two receivers. Receiver 1, which is served by AP 7, is blocked by Object 1 while Receiver 2 that is served by AP 2 is not blocked for the illustrated scenario. AP 4 provides a protection link to Receiver 1. Note that if both Receiver 1 and 2 move to the right, their blockage status can change.*

## 3. RESULTS AND DISCUSSION

### A. Single user and single AP

The results shown in Fig. 3 are based on [8] and will be used as a reference to compare with the case when resilience is implemented. Here it can be seen that depending on the parameters of the disc and the assigned AP, the percentage blockage can be between 0% and 100%.

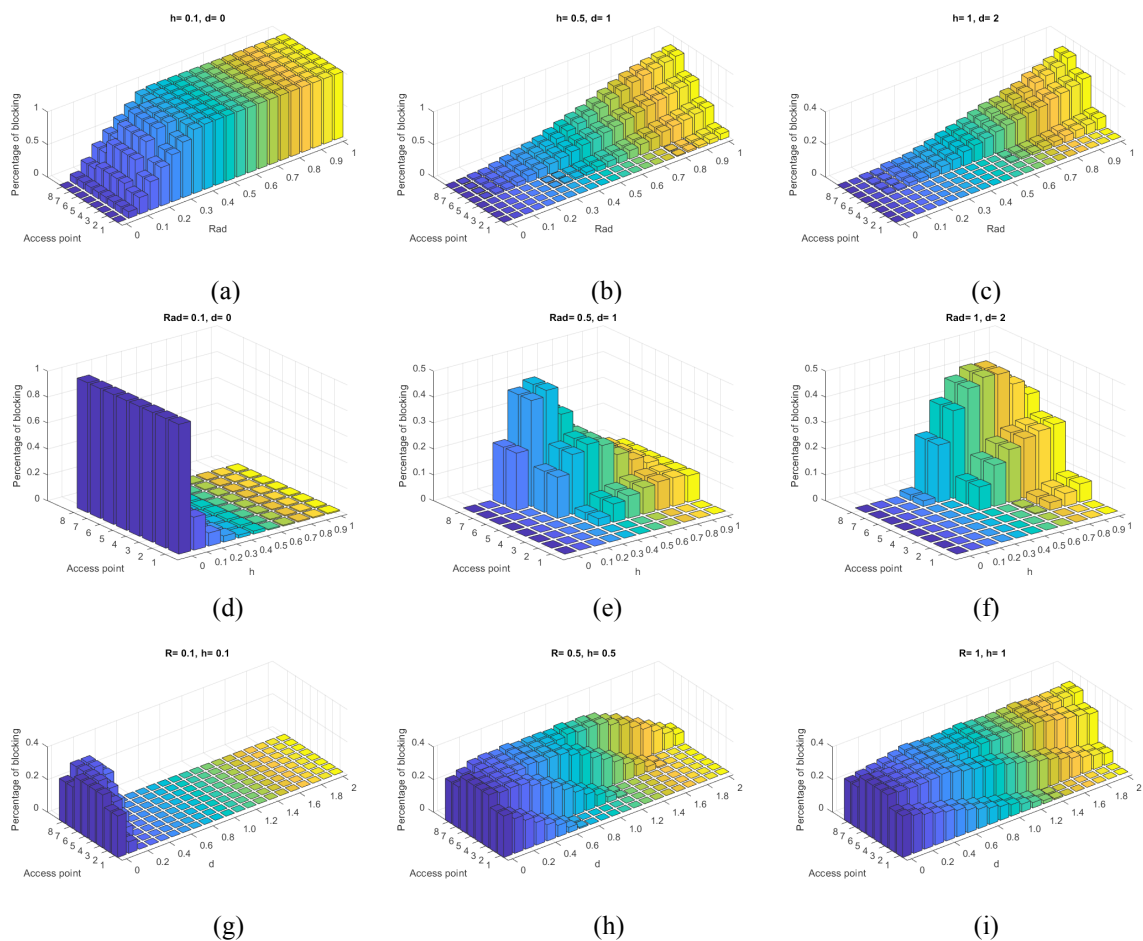

*Figure 2. Percentage Blocking when no resilience is implemented.*

## B. Resilience by assigning multiple APs

In case a user is assigned a serving link and a single or multiple protection links, and performing the same procedure as described in the previous section, it can be seen that blockage is almost mitigated as it decreases dramatically based on the combination and number of the selected APs. Figures 3, 4, and 5 show the blockage where 2, 4, and 8 access points are assigned to the user. The patterns in these figures with resilience are also similar to those in [8], but resilience helps minimize or completely avoid blockage. For these figures, scenario (a) is the worst as the obstructing object is too close with a size sufficient to cover the receiver and in this case blockage is inevitable even with maximum resilience. The first point in scenario (d) results from complete coverage of the receiver by the object and can be ignored.

In Fig. 3, where two APs are assigned to the user, the APs were selected so that there are either, (i) adjacent APs, (ii) maximally separated APs, or (iii) one AP in the middle and the other is adjacent or near the corner. Maximal separation is the best, followed by when the serving AP is to the left of the receiver as the obstructing object is assumed to be in the positive y-direction. When the APs are in the middle or to right side of the room, the blocking probability is higher.

In Fig. 4, where four APs are assigned to the user, the APs were chosen so that the APs are either towards one side of the room, towards the centre of the room, or two at the centre and the other two are towards the corner. It can also be seen that the best case is when the APs are maximally separated whether they are all towards the corner or just two are towards the corner.

Fig 5, represents maximum resilience where all APs are serving the user (in case of multiple users, time, frequency, or code multiple access techniques can be used). Here it can be seen that the link is robust and blockage goes to zero except for scenario (a) where the disc is too close to the receiver with height of 10 cm exactly above the receiver. In this case as the radius increases, the incidence angle needs to approach 90 to avoid blocking. The blockage in this case is very high and inevitable with the current system design. To avoid blockage here, cooperation with radio can be used or if optical communication is desired, angle diversity reception accompanied by relaying to other nodes at the same CF plane can be considered. The first point in scenario (d) results from complete coverage of the receiver by the object as stated before.

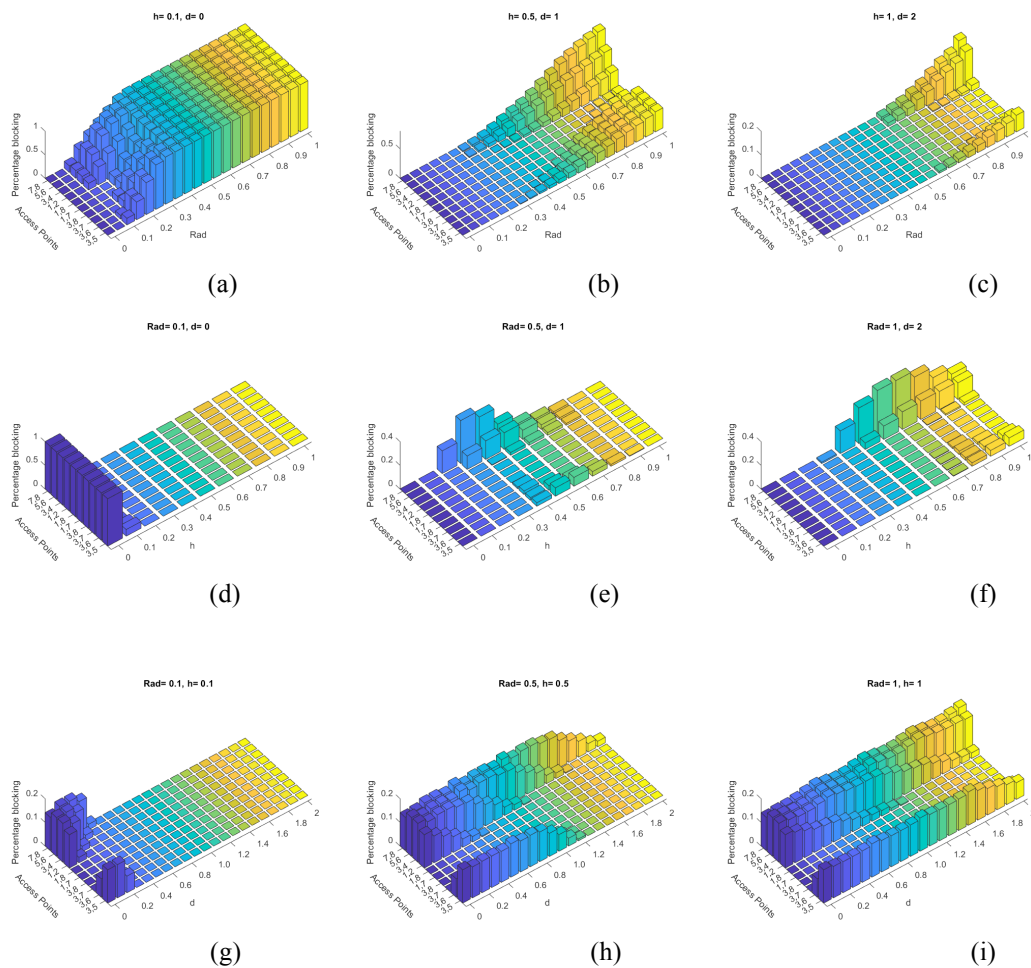

Figure 3. Percentage Blocking with one protection link in addition to the primary link.

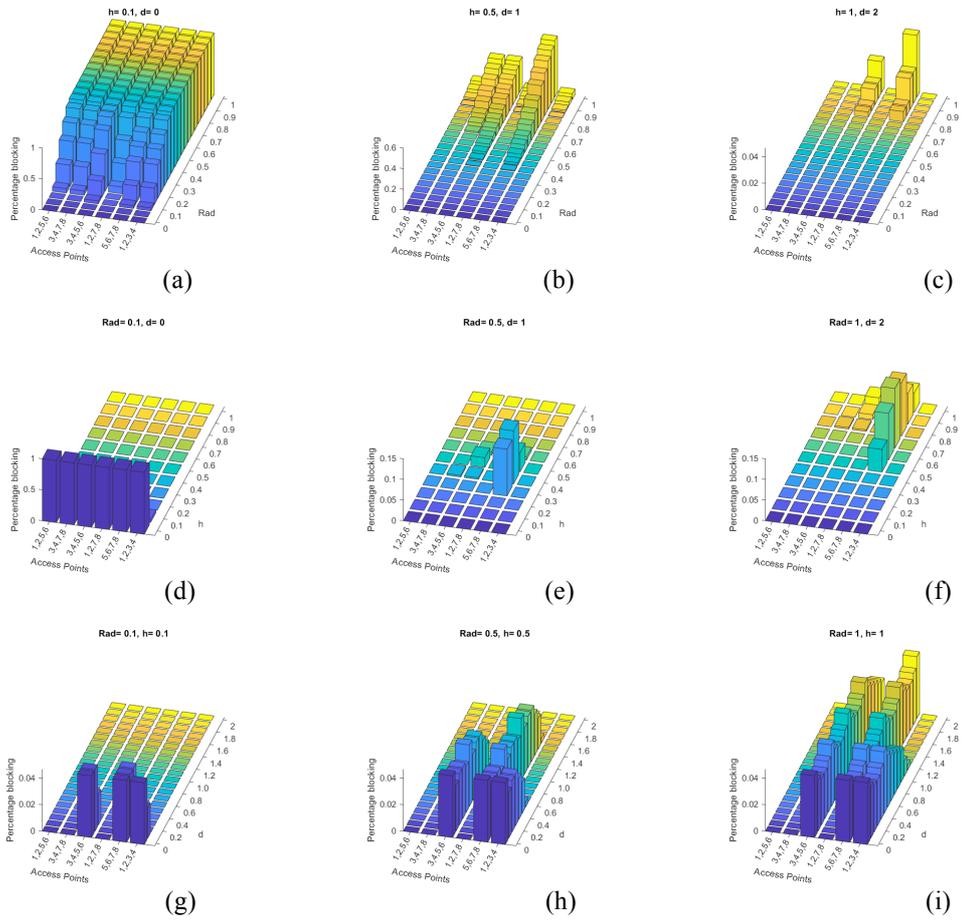

*Figure 4. Percentage Blocking with four serving APs.*

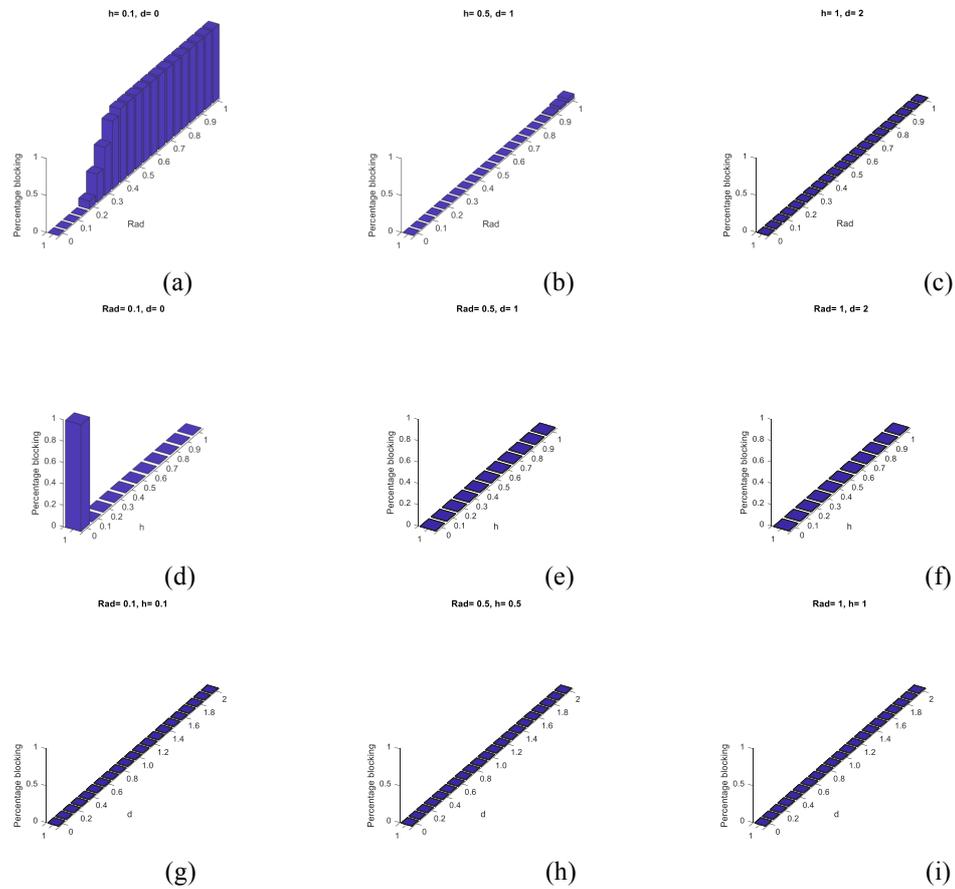

*Figure 5. Percentage of Blocking with maximum resilience.*

## 4. CONCLUSIONS AND FUTURE WORK

In this paper, the effect of direct line-of-sight beam blockage by an opaque object was studied when resilience is implemented. The blockage is studied for different values of the parameters of the obstructing object. Implementing resilience by assigning two or more links can eliminate blockage completely except for the case when the disc is too close to the receiver. In this case, either radio links can be used, or relay nodes can be employed with an optical link to other nodes on the same level as the CF. The effect of the added traffic on the network as a result of using redundant links can be studied in the future.

## ACKNOWLEDGEMENTS

The authors would like to acknowledge funding from the Engineering and Physical Sciences Research Council (EPSRC) for the TOWS project (EP/S016570/1). All data are provided in full in the results section of this paper.